\documentclass[a4paper,twoside,intlimits,10pt,pre,twocolumn]{revtex4}
\usepackage{amsmath,amsthm,amsfonts,amssymb,amsthm,dsfont,mathrsfs}
\usepackage{mathtools}
\usepackage{xspace}
\usepackage{asymptote,subfigure}
\usepackage[english]{babel}
\usepackage{pgfplots}
\pgfplotsset{compat=1.5}

\newcommand{\mE}{\mathcal{E}}
\newcommand{\mV}{\mathcal{V}}
\newcommand{\mR}{\mathcal{R}}
\newcommand{\mB}{\mathcal{B}}
\DeclareMathOperator{\Graph}{Graph}
\DeclareMathOperator{\dd}{d}		
\DeclareMathOperator{\e}{e}  
\DeclareMathOperator{\sgn}{sign}
\newcommand{\abs}[1]{{\left|#1\right|}}  
\newcommand{\Media}[1]{\left\langle #1\right\rangle}
\newcommand{\R}{\mathds{R}}                     
\newcommand{\N}{\mathds{N}}                     

\begin{document}
\title{On the one dimensional Euclidean matching problem: exact solutions, correlation functions and universality}
\author{Sergio Caracciolo}
\affiliation{Dipartimento di Fisica, University of Milan and INFN, via Celoria 16, I-20133 Milan, Italy}
\email{sergio.caracciolo@mi.infn.it}
\author{Gabriele Sicuro}
\affiliation{Dipartimento di Fisica ``E. Fermi'', University of Pisa and INFN, Largo Bruno Pontecorvo, 3, I-56127 Pisa, Italy}
\email{gabriele.sicuro@for.unipi.it}
\date{\today}
\begin{abstract}
We discuss the equivalence relation between the Euclidean bipartite matching problem on the line and on the circumference and the Brownian bridge process on the same domains. The equivalence allows us to compute the correlation function and the optimal cost of the original combinatoric problem in the thermodynamic limit; moreover, we solve also the minimax problem on the line and on the circumference. The properties of the average cost and correlation functions are discussed.
\end{abstract}
\maketitle
\section{Introduction}

Let us consider a bipartite complete graph with vertex set $\mV$ and edge set $\mE$, $\mathfrak{K}_{N,N}=\Graph(\mV,\mE)$, whose vertex set can be partitioned in two disjoint subsets of the same cardinality, $\mV=\mR\cup\mB$, $\abs{\mR}=\abs{\mB}=N\in\N$, $\mR\coloneqq\{r_1,\dots,r_N\}$, $\mB\coloneqq\{b_1,\dots,b_N\}$. Let us also introduce a weight function $w\colon\mE\to\R$, $(r_ib_j)\in\mE\mapsto w_{ij}$. In the weighted bipartite matching problem, we are interested in the permutation $\pi$ of $N$ elements such that a certain cost function
\begin{equation}
E[\pi]\coloneqq f\left(w_{1\pi(1)},\dots,w_{N\pi(N)}\right),\quad f\colon\R^N\to\R^+,
\end{equation}
is minimized. The most common form adopted in the literature for the function $f$ is simply $f(x_1,\dots,x_N)\coloneqq\sum_{i=1}^N \abs{x_i}$: in this case, the assignment problem can be solved in polynomial time using the Khun--Munkres algorithm \citep{Kuhn,Munkres1957} that, in the Edmonds and Karp's version \citep{Edmonds1972}, has $O(N^3)$ running time.

As a variation of the problem, sometimes random weights $\{w_{ij}\}$ are considered: in this case the average properties of the solution are of great interest. In the hypothesis of independently and identically distributed weights, the problem was studied using arguments borrowed both from probability theory \citep{Aldous2001} and from the theory of disordered systems \citep{Mezard1985}. Finally, if the vertices $\mV$ of the graph are associated to points randomly generated in the hypercube $\Omega=[0,1]^d$ in $d$ dimensions and $w_{ij}$ is a function of the Euclidean distance between the $r_i$ vertex and the $b_j$ vertex, the problem is called \textit{Euclidean bipartite matching problem} \citep{Mezard1988,Boniolo2012,Caracciolo2014}.

In the present paper we will consider the so-called \textit{grid-Poisson matching problem} in one dimension both with open boundary conditions (\textsc{obc}) and with periodic boundary conditions (\textsc{pbc}). The set of vertices $\mR$ is associated to a set of fixed points on the interval $\Omega\coloneqq[0,1]$ and in particular $r_i\mapsto x_i\equiv\frac{2i-1}{2N}$, $i=1,\dots, N$, whilst the set of vertices $\mB$ is associated to a set of $N$ points, $\{y_i\}_{i=1,\dots,N}$, randomly generated in the interval, such that $b_i\mapsto y_i$. We will suppose the $\mB$-vertices indexed in such a way that $i\leq j\Rightarrow y_i\leq y_j$. Finally, we will consider the following cost functional
\begin{equation}\label{costo}
E_p[\pi]=\sqrt[p]{\frac{1}{N}\sum_{i=1}^N \left[w\left(\abs{x_i-y_{\pi(i)}}\right)\right]^p},\quad p\in(1,+\infty),
\end{equation}
in which the function $w\colon[0,1]\to[0,1]$ is defined as below:
\begin{equation}
w(x)=\begin{cases}x&\text{for \textsc{obc},}\\
x\,\theta\left(\frac{1}{2}-x\right)+(1-x)\,\theta\left(x-\frac{1}{2}\right)&\text{for \textsc{pbc},}\end{cases} 
\end{equation} 
where $\theta(x)$ is the Heaviside function. In the following,
\begin{equation}
\epsilon_p\coloneqq\sqrt{N}\min_\pi E_p[\pi],
\end{equation}
and
\begin{equation}
\mathfrak e_p\coloneqq\sqrt{N^p}\min_\pi\left[ (E_p[\pi])^p\right].
\end{equation}
We will show that, for the cost functional \eqref{costo}, a solution of the problem can be obtained in the continuum limit, $N\to\infty$, not only for $p=2$ (as already shown in Ref. \citep{Boniolo2012}) but also in the $p\to\infty$ limit using well known properties of the Brownian bridge process. In fact, in the limit $p\to\infty$,
\begin{equation}
\epsilon_p\xrightarrow{p\to\infty}\sqrt{N}\min_{\pi}\max_{i}w\left(\abs{x_i-y_{\pi(i)}}\right),
\end{equation}
i.e., the problem reduces to the minimax grid-Poisson matching problem in one dimension. The problem was studied by \textcite{Leighton} for $d=2$ and \textcite{Shor1991} for $d\geq 3$. The minimax problem is related to a lot of different computational problems and the evaluation of the scaling of its cost gives directly useful informations on the computational cost of other algorithms (for a discussion of the related problems in $d=2$ see for example Ref. \citep{Leighton}). 

\section{Optimal cost and correlation function on the interval}
The solution of the grid--Poisson matching problem in one dimension for \textsc{obc} is easily found by simple arguments \citep{Boniolo2012} for $p>1$ and cost functional \eqref{costo}: in this case, in fact, the optimal matching is always \textit{ordered}, i.e. $\pi(i)=i$, in the hypothesis above that $i<j\Rightarrow y_i\leq y_j$. Note that, due to the monotony of the function $x\mapsto x^p$, the optimal solution for the cost $E_p$ is also optimal for the cost $(E_p)^p$. The probability density distribution for the position of the $i$-th $\mB$-point is:
\begin{equation}
\Pr\left(y_i\in \dd y\right)=\binom{N}{i}y^i(1-y)^{N-i}\frac{i}{y}\dd y,
\end{equation}
where we used the short-hand notation $x\in \dd z\Leftrightarrow x\in[z,z+\dd z]$. In the $N\to\infty$ limit, a non trivial result is obtained introducing the variable $m(y)$
\begin{equation}
m(y)\coloneqq \sqrt{N}M(y),\quad M(y)\coloneqq \frac{i}{N}-y
\end{equation}
expressing the rescaled (signed) distance between a $\mB$-point in $[y,y+\dd y]$ and its corresponding $\mR$-point in the optimal matching. We finally obtain a distribution for the variable $m(y)$ depending on the position on the interval $y\in[0,1]$:
\begin{equation}
\Pr\left(m(y)\in\dd m\right)=\frac{\e^{-\frac{m^2}{2 y(1-y)}}}{\sqrt{2\pi y(1-y)}}\dd m.\label{bb}
\end{equation}
The distribution \eqref{bb} is the one of a Brownian bridge on the interval $[0,1]$, a continuous time stochastic process defined as
\begin{equation}\mathsf B(t)\coloneqq\mathsf W(t)-t\mathsf W(1),\quad t\in[0,1],\end{equation}
where $\mathsf W(t)$ is a Wiener process. The joint distribution of the process can be derived similarly (see Appendix). In particular, the covariance matrix for the $2$-points joint distribution has the form, for $s,t\in[0,1]$ (see eq.~\eqref{JBB2}),
\begin{equation}\label{covm}
\Sigma_2=\begin{pmatrix}
2\phi(s)&&\phi(s)+\phi(t)-\phi(t-s)\\\phi(s)+\phi(t)-\phi(t-s)&&2\phi(t)
\end{pmatrix},
\end{equation}
where we introduced the function
\begin{equation}
\phi(x)\coloneqq \abs{x}\frac{1-\abs{x}}{2}.
\end{equation}
Averaging over the positions $s,t$ and fixing the distance $\tau\coloneqq\abs{t-s}$, we have
\begin{equation}
\bar \Sigma_2(\tau)=\begin{pmatrix}
c&&c-\phi(\tau)\\c-\phi(\tau)&&c
\end{pmatrix},\quad c=\frac{1}{6}.
\end{equation}

The Euclidean matching problem on the interval $[0,1]$ with open boundary conditions and cost functional \eqref{costo} is therefore related to the Brownian bridge in the $N\to\infty$ limit. By using this correspondence, the correlation function $\forall p>1$ is computed as
\begin{equation}\label{cfobc}
\overline{\Media{m(t)m(t+\tau)}}=\frac{1}{6}-\phi(\tau).
\end{equation}
where the average $\overline{\bullet}$ is intended on the position $t$, whilst we denoted by $\Media{\bullet}$ the average over different realisations of the problem. This theoretical prediction was confirmed numerically. Introducing the normalised variable
\begin{equation}
\sigma(t)=\frac{m(t)}{\abs{m(t)}}=\sgn(m(t)),
\end{equation}
\textcite{Boniolo2012} computed also the correlation function for this quantity, finding
\begin{multline}
\Media{\sigma(s)\sigma(t)}=\frac{2}{\pi}\arctan\sqrt{\frac{\min(s,t)(1-\max(s,t))}{\abs{t-s}}}\\
\Rightarrow\int_0^{1-t}\Media{\sigma(s)\sigma(s+t)}\dd s=\frac{1-\sqrt{t}}{1+\sqrt{t}}.
\end{multline}
Both formulas were confirmed numerically. Note that all the results above holds $\forall p>1$ in the case of open boundary conditions.

Let us now compute the average cost of the matching. From eq.~\eqref{bb} we obtained that
\begin{equation}
\Media{\mathfrak e_p}\xrightarrow{N\to\infty}\int_0^1\Media{\abs{\mathsf B(t)}^p}\dd t=\frac{1}{2^\frac{p}{2}(p+1)}\Gamma\left(\frac{p}{2}+1\right).
\end{equation}

\begin{figure}
\includegraphics{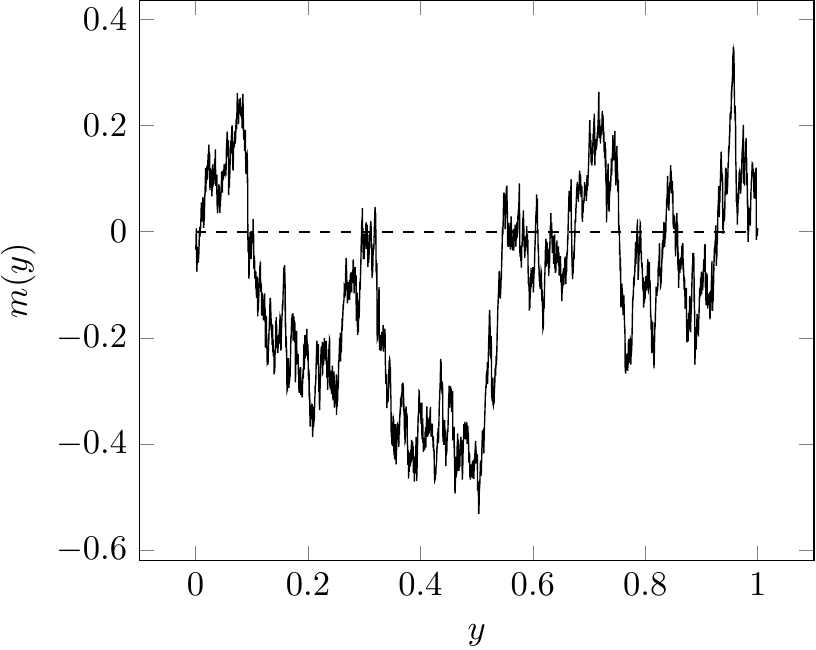}
\caption{Plot of $m(y_i)=\sqrt{N}\left(\frac{i}{N}-y_i\right)$ for a certain realisation of the problem with $N=4096$.}
\end{figure}

\begin{figure}
\includegraphics{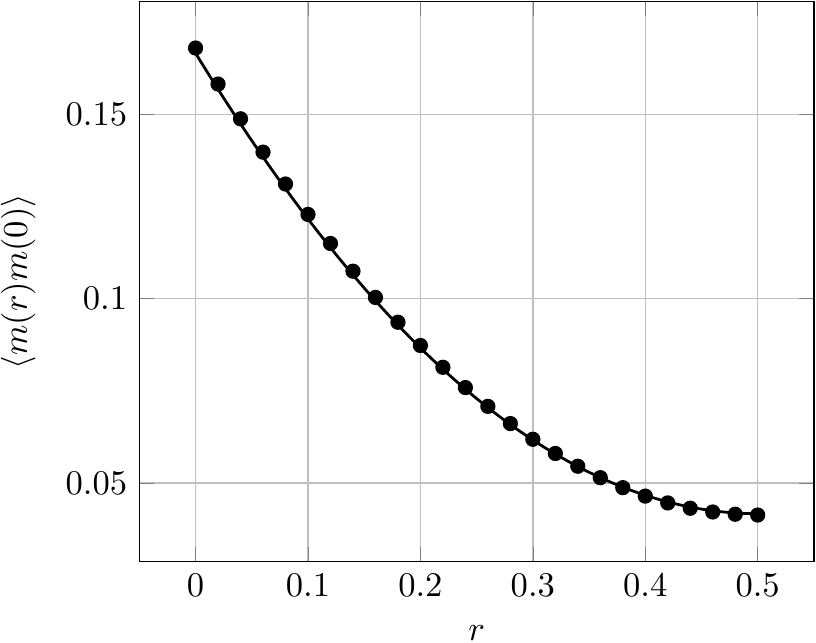}
\caption{Correlation function $\forall p>1$ in the \textsc{obc} case for $N=1000$ obtained averaging over $10^4$ realisations (for clarity, not all data are represented).}\label{fcfobc}
\end{figure}

\begin{figure}
\includegraphics{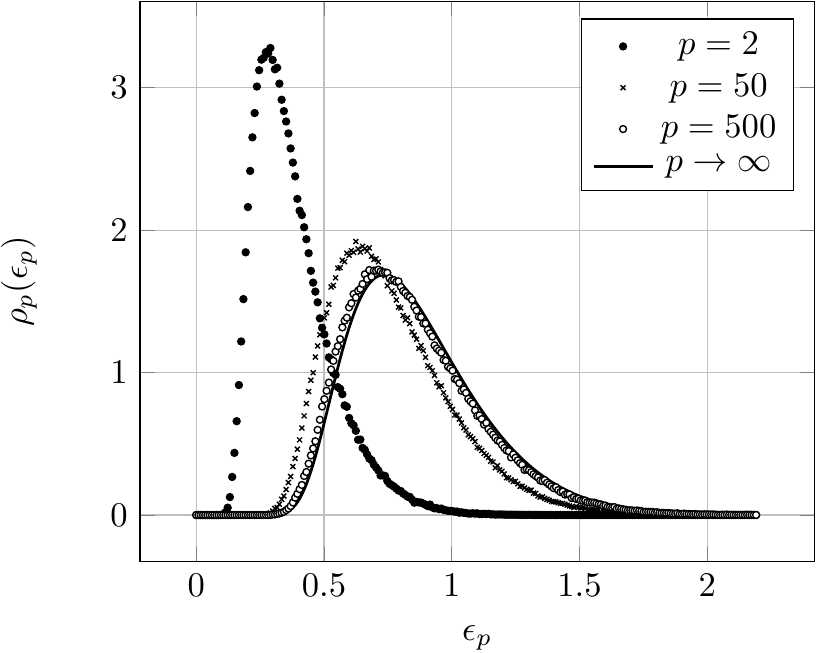}
\caption{Distribution density of the rescaled cost for different values of $p$ computed with $5\cdot 10^5$ iterations and $N=5000$ for the \textsc{obc} case.}\label{fcobcD}
\end{figure}

Moreover, the optimal cost $\epsilon_p$ in the $N\to\infty$ limit can be written as
\begin{equation}
\Media{\epsilon_p}\xrightarrow{N\to\infty}\Media{\left(\int_0^1\abs{\mathsf B(t)}^p\dd t\right)^{\frac{1}{p}}}.
\end{equation}
Although the previous expression is difficult to evaluate exactly for finite $p$ (see for example Ref. \citep{Shepp10} for additional information about the distribution of $\mathsf X_p\coloneqq \int_0^1\abs{\mathsf B(t)}^p\dd t$), the calculation can be easily performed in the relevant limit $p\to\infty$, being
\begin{equation}
\Media{\left(\int_0^1\abs{\mathsf B(t)}^p\dd t\right)^{\frac{1}{p}}}\xrightarrow{p\to\infty}\Media{\sup_{t\in[0,1]}\abs{\mathsf B(t)}}.
\end{equation}
The distribution of the supremum of the absolute value of the Brownian bridge is the well known Kolmogorov distribution \citep{dudley2002real}
\begin{equation}\label{kolmogorov}
\Pr\left(\sup_{t\in[0,1]}\abs{\mathsf B(t)}<u\right)=\sum_{k=-\infty}^{+\infty}(-1)^{k}\e^{-2k^2u^2}
\end{equation}
and therefore
\begin{equation}
\Media{\epsilon_p}\xrightarrow{N\to\infty}\sqrt{\frac{\pi}{2}}\ln 2.
\end{equation}
In figure \ref{fcobcD} we plotted $\rho_p(u)\coloneqq\frac{\dd}{\dd u}\left[\Pr\left(\epsilon_p\leq u\right)\right]$ for different values of $p$: observe that $\rho_p$ approaches the Kolmogorov distribution in the large $p$ limit.

Finally, observe also that the variance of $\mathfrak{e}_p$
\begin{equation}
\Media{\left(\mathfrak{e}_p-\Media{\mathfrak{e}_p}\right)^2}=\frac{(1+p)\Gamma(p+2)-(2p+1)\Gamma^2\left(\frac{p}{2}+1\right)}{2^p(p+1)^2(2p+1)}
\end{equation}
increases with $p$ and that
\begin{equation}
\frac{\Media{\left(\mathfrak{e}_p-\Media{\mathfrak{e}_p}\right)^2}}{\Media{\mathfrak{e}_p}^2}=\frac{(p+1) \Gamma (p+2)}{(2 p+1) \Gamma \left(\frac{p}{2}+1\right)^2}-1\xrightarrow{p\to\infty}+\infty.
\end{equation}
From a numerical point of view, this means that a computation of $\Media{\mathfrak{e}_p}$ requires a larger amount of iterations as $p$ increases and fluctuations around the mean value become extremely large in the $p\to\infty$ limit. On the other hand, fluctuations of the optimal cost $\epsilon_p$ around its mean value $\Media{\epsilon_p}$ for $p\to\infty$ remain finite
\begin{equation}
\Media{\left(\epsilon_p-\Media{\epsilon_p}\right)^2}\xrightarrow{p\to\infty}\frac{\pi^2}{12}-\frac{\pi}{2}\ln^2 2.
\end{equation}
This fact allows to perform a precise computation of $\Media{\epsilon_p}$ for large $p$.

\section{Optimal cost and correlation function on the circumference}\label{cerchio}
Let us now consider the case of periodic boundary conditions. As discussed in Ref. \citep{Boniolo2012}, the solution for both the cost $E_p$ and the cost $(E_p)^p$, with $p\in(1,+\infty)$, is again ordered; however, in this case the mapping is $i\mapsto \pi(i)=i+\lambda\mod N$ for a certain $\lambda\in\{0,1,\dots,N-1\}$. In the continuum limit, the solution is a generalised Brownian bridge, $m_p(t)=\mathsf B(t)+\lambda_p$, $t\in[0,1]$, for a certain constant $\lambda_p\in\R$ depending on $p$. The constant $\lambda_p$ can be found by optimality condition on the cost functional \eqref{costo}:
\begin{equation}
\frac{\partial}{\partial\lambda_p}\left(\int_0^1\abs{\mathsf B(t)+\lambda_p}^p\dd t\right)^\frac{1}{p}=0.
\end{equation} 
The previous equation can be solved only for $p=2$ and $p\to+\infty$. For $p=2$, $\lambda_2=-\int_0^1\mathsf{B}(\tau)\dd \tau$; therefore, $m_2(t)=\mathsf B(t)-\int_0^1\mathsf{B}(t)\dd t$ and
\begin{equation}
\overline{\Media{m_2(t)m_2(t+\tau)}}=\frac{1}{12}-\phi(\tau).
\end{equation}
For $p\to\infty$, $\left(\int_0^1\abs{\mathsf B(t)+\lambda_p}^p\dd t\right)^\frac{1}{p}\xrightarrow{p\to\infty}\sup_{t\in[0,1]}\abs{\mathsf B(t)+\lambda_\infty}$ and therefore the optimality condition becomes
\begin{multline}\label{lambdainf}
\frac{\partial}{\partial\lambda}\left(\sup_{t\in[0,1]}\abs{\mathsf B(t)+\lambda}\right)=0\\
\Rightarrow \lambda_\infty=-\frac{\sup_{t\in[0,1]}\mathsf B(t)+\inf_{t\in[0,1]}\mathsf B(t)}{2}.
\end{multline}
Indeed, we have that
\begin{multline}\textstyle
\sup_t\abs{\mathsf B(t)+\lambda}=\\\textstyle=\abs{\sup_t\mathsf B(t)+\lambda}\theta(\abs{\sup_t\mathsf B(t)+\lambda}-\abs{\inf_t\mathsf B(t)+\lambda}) \\\textstyle+\abs{\inf_t\mathsf B(t)+\lambda}\theta(\abs{\inf_t\mathsf B(t)+\lambda}-\abs{\sup_t\mathsf B(t)+\lambda})\end{multline}
from which the eq.~\eqref{lambdainf} is derived. We have therefore, for $t\in[0,1]$,
\begin{equation}\label{minf}
m_\infty(t)=\mathsf B(t)-\frac{\sup_{s\in[0,1]}\mathsf B(s)+\inf_{s\in[0,1]}\mathsf B(s)}{2}.
\end{equation}

\begin{figure}\label{fcfpbc}
\includegraphics{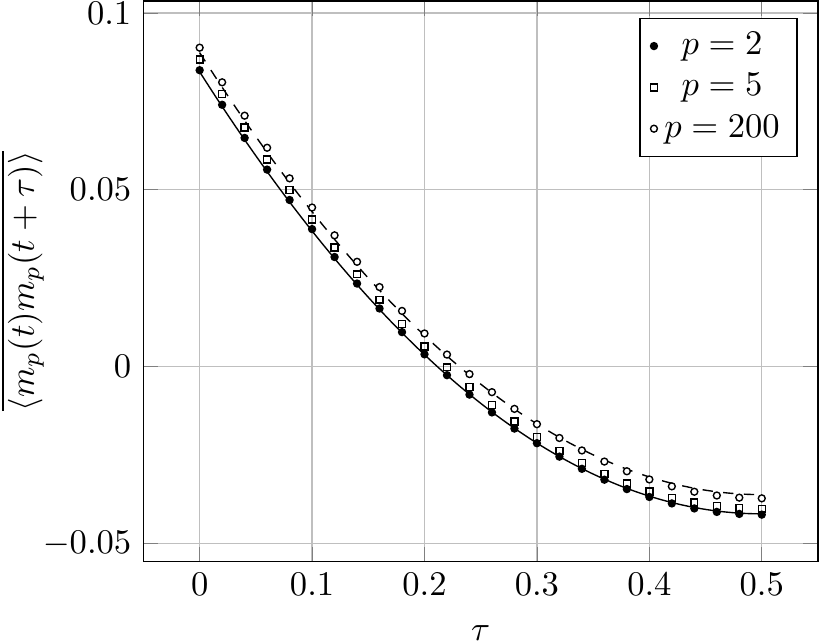}
\caption{Correlation function in the \textsc{pbc} case for $N=1000$ obtained averaging over $10^4$ realisations (for clarity, not all data are represented). The continuous line is the theoretical prediction for $p=2$; the dashed line corresponds to the $p\to\infty$ case.}
\end{figure}

\begin{figure}
\includegraphics{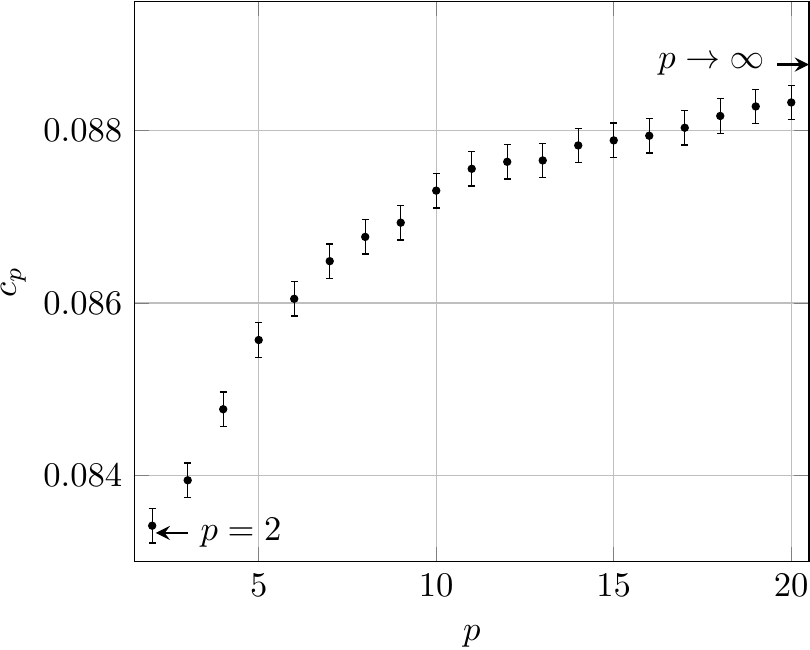}
\caption{Value of $c_p\coloneqq\overline{\Media{m_p^2(t)}}$ in the \textsc{pbc} case for $N=1000$ obtained from the fit of the correlation function, averaging over $5000$ realisations, for different values of $p$.}\label{fcf0pbc}
\end{figure}

\begin{figure}
\includegraphics{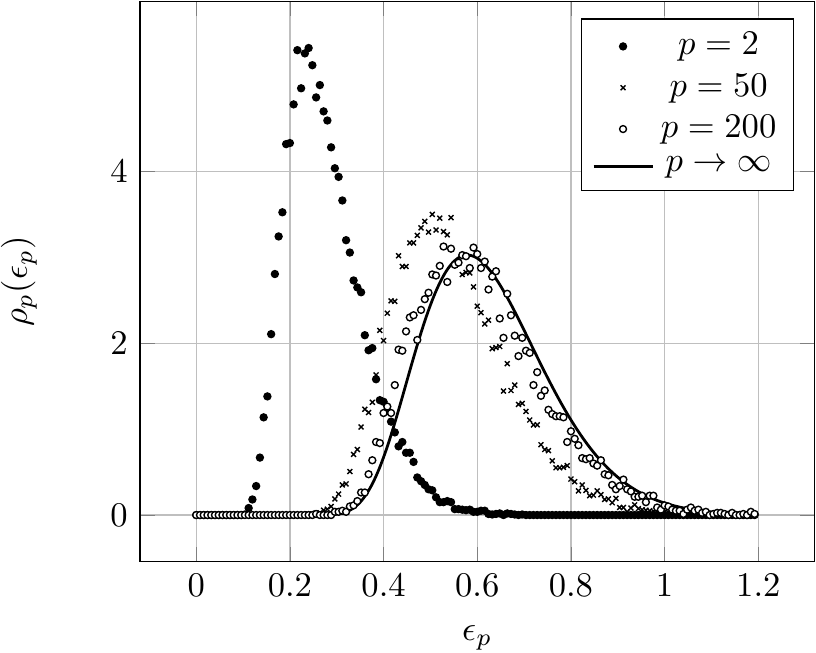}
\caption{Density distribution of the rescaled cost for different values of $p$ computed with $2\cdot 10^4$ iterations and $N=1000$ for the \textsc{pbc} case.}\label{fcpbcD}
\end{figure}

\begin{figure}
\includegraphics{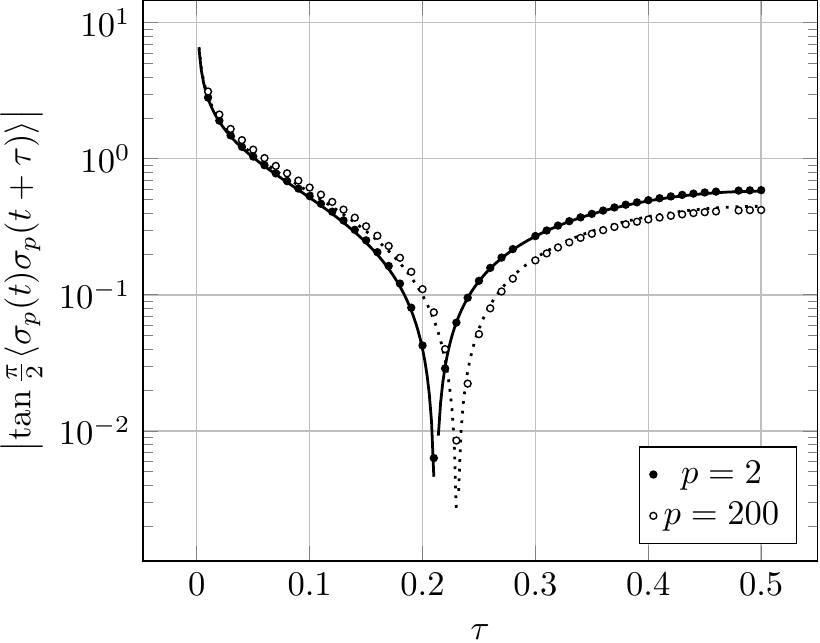}
\caption{Plot for the normalised map correlation function \eqref{ss} obtained with $10^4$ iterations and $N=1000$ for the \textsc{pbc} case. The continuous line is the theoretical prediction for $p=2$, whilst the dotted line is obtained from the theoretical prediction for $p\to\infty$ The pole is at $\tau(p)=\frac{1}{2}-\frac{\sqrt{1-8c_p}}{2}$.}\label{fcpbcSS}
\end{figure}

The correlation function can be directly found using the known joint distributions for the Brownian bridge and its sup, eq.~\eqref{bbsup}, and for the sup and inf of a Brownian bridge, eq.~\eqref{bbjmb}. After some calculations we obtain
\begin{equation}\label{covinf}
\overline{\Media{m_\infty(t)m_\infty(t+\tau)}}=\frac{2-\zeta(2)}{4}-\phi(\tau),
\end{equation}
where $\zeta(2)\coloneqq\sum_{n=1}^\infty\frac{1}{n^2}=\frac{\pi^2}{6}$. The value $\overline{\Media{m^2_\infty(t)}}=\frac{2-\zeta(2)}{4}=0.0887665\dots$ is very close to the value obtained for $p=2$, $\overline{\Media{m^2_2(t)}}=\frac{1}{12}=0.08\bar 3$. In figure \ref{fcf0pbc} we plot the values of $\overline{\Media{m^2_p(t)}}$ as function of $p$. Note, finally, that due to the fact that we have imposed \textsc{pbc}, $\overline{\Media{\bullet}}\equiv\Media{\bullet}$ holds in all the previous formulas.

Let us now introduce the normalised transport field
\begin{equation}
\sigma_p(t)\coloneqq\frac{m_p(t)}{\abs{m_p(t)}}=\sgn(m_p(t)).
\end{equation}
The correlation function $\Media{\sigma_p(s)\sigma_p(t)}$ can be computed from the covariance matrix observing that the process is still Gaussian. The correlation function is found in the form
\begin{equation}\label{ss}
\Media{\sigma_p(s)\sigma_p(t)}=\frac{2}{\pi}\arctan\frac{c_p-\phi(t-s)}{\sqrt{\phi(t-s)\left(2c_p-\phi(t-s)\right)}}
\end{equation}
where $c_2=\frac{1}{12}$ for $p=2$ and $c_\infty=\frac{2-\zeta(2)}{4}$.

The optimal cost in the $p\to\infty$ limit can be evaluated as the average spread of the Brownian bridge. Denoting 
\begin{equation}
\xi\coloneqq \sup_{s\in[0,1]}\mathsf B(s)-\inf_{s\in[0,1]}\mathsf B(s),
\end{equation} 
the distribution of the spread $\xi$ is given by \citep{dudley2002real}
\begin{equation}\label{spread}
\Pr\left(\xi<u\right)
=\theta_3\left(\e^{-2 u^2}\right)+u\frac{\dd}{\dd u}\theta_3\left(\e^{-2 u^2}\right),
\end{equation}
where
\begin{equation}
\theta_3(x)\equiv\vartheta_3(0,x),\quad \vartheta_3(z,q)\coloneqq 1+2\sum_{n=1}^\infty q^{n^2}\cos(2nz)\end{equation} 
is the third Jacobi theta function. From eq.~\eqref{spread} the distribution of the optimal cost in the $p\to\infty$ limit is easily derived. Moreover,
\begin{multline}\label{mepbc}
\Media{\epsilon_p}\xrightarrow[p\to\infty]{N\to\infty}\Media{\epsilon_\infty}=\frac{\Media{\xi}}{2}=\\=\frac{1}{2}\left(\Media{\sup_{s\in[0,1]}\mathsf B(s)}-\Media{\inf_{s\in[0,1]}\mathsf B(s)}\right)=\frac{1}{2}\sqrt{\frac{\pi}{2}}
\end{multline}
with corresponding variance
\begin{equation}
\Media{\left(\epsilon_p-\Media{\epsilon_p}\right)^2}\xrightarrow{p\to\infty}\frac{\pi^2-3\pi}{24}.
\end{equation}
\section{General solution: a continuum approach}

In the present section we will justify and generalise the previous results looking at a continuum version of the problem, the so called \textit{Monge--Kantorovi\v{c} problem}.

Let us consider the interval $\Omega\coloneqq [0,L]\subset\R$, $L\in\R^+$ and let us suppose also that two different measures are given on $\Omega$, i.e., the uniform (Lebesgue) measure
\begin{equation}
\dd m(x)\coloneqq\frac{1}{L}\dd x,
\end{equation}
and a non uniform measure $\dd n(x)$ with measure density $\nu(x)$,
\begin{multline}\label{MK}\textstyle
\dd n(x)\coloneqq \nu(x)\dd x=\frac{\dd x}{L}+\\\textstyle \dd x\sum_{k=1}^\infty\left( \nu_1(k)\cos\frac{2\pi k x}{L}+\nu_2(k)\sin\frac{2\pi k x}{L}\right).
\end{multline}
We ask for the optimal map $\mu(x)\colon\Omega\to\Omega$ such that the following transport condition is satisfied
\begin{equation}\label{admcond}
\frac{1}{L}\int_A\dd x=\int_{\mu^{-1}(A)}\dd n(x)\quad\forall A\subset\Omega\text{ measurable.}
\end{equation}
and $\mu$ minimises the following functional
\begin{equation}\label{funcost}
\mathfrak{E}_p[\mu]\coloneqq\int_0^L\abs{x-\mu(x)}^p\dd n(x),\quad p\in\R^+.
\end{equation}
It can be proved \citep{villani2008optimal} that for $p>1$ eq.~\eqref{admcond} can be rewritten as a change-of-variable formula:
\begin{equation}\label{mk}
L\dd n(x)=\dd\mu(x).
\end{equation}
We will restrict therefore to the case $p>1$. Imposing \textsc{pbc}, that is $\mu_p(0)=\mu_p(L)-L$, the solution of \eqref{mk} determines the optimal map up to a constant $\mu_p(0)$ as
\begin{equation}\textstyle
\mu_p(x)=x+\mu_p(0)+\Phi(x).
\end{equation}
In the previous equation we have introduced $\Phi(x)$
\begin{multline}\label{Phi}
\textstyle\Phi(x)\coloneqq \sum_{k=1}^\infty\nu_1(k)\frac{L^2\sin\left(\frac{k\pi x}{L}\right)}{\pi k}\cos\left(\frac{k\pi x}{L}\right)\\\textstyle+\sum_{k=1}^\infty\nu_2(k)\frac{L^2}{\pi k}\sin^2\left(\frac{k\pi x}{L}\right).
\end{multline}
Note that $\Phi(0)=\Phi(L)=0$. The value of $\mu_p(0)$ must be determined requiring that the functional \eqref{funcost} is minimum: we have that
\begin{multline}\label{mu0}\textstyle
\frac{\partial}{\partial\mu_p(0)}\int_0^L\abs{\mu_p(0)+\Phi(x)}^p\dd n(x)=\\\textstyle=p\int_0^L\sgn\left(\mu_p(0)+\Phi(x)\right)\abs{\mu_p(0)+\Phi(x)}^{p-1}\dd n(x)=0.
\end{multline}
If instead \textsc{obc} are considered, then $\mu_p(0)=0$ and the solution is obtained explicitly $\forall p>1$.

Let us now suppose that $L\equiv N\in\N$ and that the measure $n(\dd x)$ is obtained as a limit measure of a random atomic measure of the form
\begin{multline}\label{nN}\textstyle
\dd n_N(x)\coloneqq\frac{\dd x}{N}\sum_{i=1}^N\delta\left(x-\eta_i\right)\\\textstyle
=\dd\left(\frac{1}{N}\sum_{i=1}^N\theta\left(x-\eta_i\right)\right),
\end{multline}
where $\{\eta_i\}_{i=1,\dots,N}$ is a set of $N$ points uniformly randomly distributed in $\Omega$. The previous measure can be written as
\begin{multline}\label{nserie}\textstyle
n_N(x)=\sum_{k=1}^\infty\sqrt{\frac{2}{N}}\frac{Z_k}{\pi k}\sin\frac{\pi k x}{N}+\frac{x}{N}\\
\textstyle+\frac{1}{N}\sum_{i=1}^N\eta_i-1.
\end{multline}
where we have introduced
\begin{multline}
Z_k\equiv Z_k(x)\coloneqq\frac{1}{N}\sum_{i=1}^Nz_i(x),\\ z_i(x)\coloneqq-\sqrt{2N}\cos \left(k\pi\frac{2\eta_i+x}{N}\right).
\end{multline}
Observe now that $Z_k(x)$ is a sum of independent identically distributed random variables. From the central limit theorem, we have that $Z_k(x)$ is normally distributed as
\begin{equation}
Z_k\sim\mathcal N\left(0,1\right)\quad \forall k\in\N.
\end{equation}
Remarkably the previous distribution does not depend on $x$. Moreover, the $Z_k$ and $Z_l$ are independent random variables for $k\neq l$, being Gaussian distributed and $\Media{Z_lZ_k}=0$, where the average $\Media{\bullet}$ is intended over the possible values $\{\eta_i\}_i$. In eq.~\eqref{nserie} the Karhunen--Lo\`eve expansion for the Brownian bridge \citep{Barczy2010} on the interval $[0,N]$ appears:
\begin{equation}\label{kl}
\mathsf B_N(x)\coloneqq \sum_{k=1}^\infty\sqrt{2N}\frac{Z_k}{\pi k}\sin\frac{\pi k x}{N},\quad Z_k\sim\mathcal N\left(0,1\right)\quad \forall k\in\N.
\end{equation} It follows that $n_N(x)$ can be written for large $N$, up to irrelevant additive constants, as
\begin{equation}
n_N(x)\simeq\frac{1}{N}\mathsf B_N(x)+\frac{x}{N}
\end{equation}
and therefore we cannot associate a density measure to it, due to the fact that $n_N(x)$ is not differentiable.

However the solution of the matching problem in the continuum can be still obtained directly from eq.~\eqref{MK}; considering \textsc{pbc}, then
\begin{equation}
\mu_p(x)=\mu_p(0)+x+\mathsf B_N(x),\quad x\in[0,N].
\end{equation}
Denoting by
\begin{equation}
M_p(x)\coloneqq\mu_p(x)-x,
\end{equation}
it follows that $\forall p>1$
\begin{equation}
\Media{M_p(x)M_p(y)}=c_p(N)-N\phi\left(\frac{x-y}{N}\right).\label{cfcont}
\end{equation}
where $c_p(N)$ is a constant depending on $N$ and $p$. Adopting the notation
\begin{equation}
F(N)\sim G(N)\Leftrightarrow 0<\lim_{N\to\infty}\frac{F(N)}{G(N)}<+\infty,
\end{equation}
for two positive real functions $F$ and $G$ depending on $N$, note that 
\begin{equation}
c_p(N)\sim c_pN
\end{equation}
for some positive constant $c_p$ depending on $p$. Indeed, in the discrete case, from the fact that the solution must be ordered, in the large $N$ limit it can be easily seen that
\begin{equation}
\min_\mu\mathfrak{E}_p[\mu]\sim N^{\frac{p}{2}},
\end{equation}
where $\mathfrak{E}_p$ is the cost functional \eqref{funcost} in which the measure \eqref{nN} is adopted. Therefore, $\mu_p(0)=M_p(0)\sim M_p(x)\sim\sqrt N$. Moreover, note also that $\overline{\Media{M^2(t)}}< N^2\Rightarrow c_p(0)=0$. For $p=2$, eq.~\eqref{mu0} becomes
\begin{multline}\textstyle
\mu_2(0)=-\frac{1}{N}\int_0^{N}\mathsf B_N(x)\circ\dd\mathsf B_N(x)-\frac{1}{N}\int_0^{N}\mathsf B_N(x)\dd x\\
\textstyle=-\frac{1}{N}\int_0^{N}\mathsf B_N(x)\dd x
\end{multline}
where the first integral is intended in the Stratonovi\v{c} sense. The result is in agreement with the one presented in Section \ref{cerchio}. If we consider the transport cost functional
\begin{equation}\label{funcost2}
E_p[\tilde \mu]\coloneqq\sqrt[p]{\int_0^N\abs{x-\tilde \mu(x)}^p\dd n(x)}
\end{equation}
the matching problem has the same solution obtained for the cost \eqref{funcost} for all finite values of $p$, $\mu_p(x)\equiv\tilde\mu_p(x)\Rightarrow\mu_p(0)\equiv\tilde\mu_p(0)$, due to the fact that the function $f(x)=\sqrt[p]{x}$ is monotone. However, for the functional cost \eqref{funcost2}, in the $p\to\infty$ limit, we can reproduce the computations of Section \ref{cerchio}, obtaining
\begin{equation}
\lim_{p\to\infty}\mu_p(0)=-\frac{\sup_{s\in[0,N]} \mathsf B_N(s)+\inf_{s\in[0,N]} \mathsf B_N(s)}{2}.
\end{equation}
If \textsc{obc} are considered, then $\mu_p(0)=0$ and we have simply, $\forall p>1$,
\begin{equation}
M_p(x)\equiv M(x)=\mathsf B_N(x),\quad x\in[0,N];
\end{equation}
It can be easily seen that
\begin{equation}
\frac{1}{N}\int_0^N\Media{M(x)M(x+r)}\dd x=c(N)+N\phi\left(\frac{r}{N}\right)
\end{equation}
where $c(N)=\frac{N}{6}$.

\section{Conclusions}
In the present work we solved the Euclidean bipartite matching problem on the interval $[0,1]$, using the cost functional \eqref{costo} in the continuum limit, $N\to\infty$, for $p\to\infty$, both with open boundary conditions and with periodic boundary conditions. The solution is based on the exact correspondence between the optimal map and a Brownian bridge process on the same interval. Moreover, we computed the correlation function for the optimal map and we observed that in all considered cases it has the form
\begin{equation}
\overline{\Media{m_p(t)m_p(t+\tau)}}=c_p-\phi(\tau),\quad \tau\in[0,1],
\end{equation}
for some constant $c_p$ depending on the adopted boundary conditions and on the value $p$ for the optimal cost. Note that if we consider the problem of a matching of $N$ random $\mB$-points to $N$ lattice $\mR$-points on the interval $[0,N]$, for $s,t\in[0,N]$ the correlation function assumes the form
\begin{equation}
G(\tau)\coloneqq \overline{\Media{M_p(t)M_p(t+\tau)}}=c_pN-\frac{\abs{\tau}}{2}\left(1-\frac{\abs{\tau}}{N}\right),\label{G}
\end{equation}
where $M_p(t)$ is the signed distance between $t\in[0,N]$ and its corresponding inverse image under the action of the optimal map. It follows that for $N\to\infty$ the correlation function $G(\tau)$ has a divergent part, $G(0)= c_pN$, depending through $c_p$ on the specific details of the problem (e.g., the boundary conditions adopted or the value of $p$), a universal finite part $-\frac{\abs{\tau}}{2}$ and a (universal) finite size correction $\frac{\tau^2}{2N}$. We obtained also numerical evidences of the validity of eq.~\eqref{G} for different values of $p$ both with \textsc{obc} and with \textsc{pbc}: an exact derivation of $c_p$ for \textsc{obc} $\forall p>1$ and for $c_2$ and $c_\infty$ for \textsc{pbc} was presented. This fact suggests that all Euclidean matching problems in one dimension with strictly convex cost functionals belong to the same universality class and that the specific details of the model determine only the value of the constant $c_p$ in the divergent contribution $G(0)$. Similarly, on the interval $[0,N]$ eq.~\eqref{ss} becomes
\begin{multline}
\Media{\sigma_p(t)\sigma_p(t+\tau)}=\\
=\frac{2}{\pi}\arctan\frac{2c_pN-\frac{\tau}{2}\left(1-\frac{\tau}{N}\right)}{\sqrt{\frac{\tau}{2}\left(1-\frac{\tau}{N}\right)\left(4c_pN-\tau\left(1-\frac{\tau}{N}\right)\right)}}\\
= 1-\frac{1}{\pi}\sqrt{\frac{2\tau}{c_p N}}+o\left(\frac{1}{\sqrt{N}}\right),
\end{multline}
in which the universal part is given by the constant $1$ and finite size corrections scale as $\frac{1}{\sqrt{N}}$ up to a scaling factor depending on $c_p$.
\section{Acknowledgements}
The authors are grateful to Luigi Ambrosio, from Scuola Normale Superiore in Pisa, and Andrea Sportiello, from Universit\'e Paris 13, for useful and stimulating discussions.

\appendix
\section{Joint distribution for the solution of the problem with open boundary conditions}
In the present section we derive the joint distribution of the matching map $m$ for the solution of the Euclidean bipartite matching problem on the line. In the hypothesis that $0< t_1< t_2<\dots< t_K< 1$, $K<N$, let us consider $K$ $\mB$-points $\{y_{i_k}\}_{k=1,\dots,K}$, $i_1< i_2<\dots<i_K$, and evaluate the following quantity:
\begin{multline}
\Pr\left(\left\{y_{i_k}\in \dd t_k\right\}_{k=1,\dots,K}\right)=\\
=\frac{N!(1-t_K)^{N-i_K}}{(N-i_K)!}\prod_{k=1}^K\frac{(\Delta t_k)^{\Delta i_k-1}\dd t_k}{(\Delta i_k-1)!}\\
=\binom{N}{\Delta i_1 \dots \Delta i_{K+1}}\prod_{k=1}^{K+1}(\Delta t_k)^{\Delta i_k}\prod_{k=1}^K\frac{\Delta i_k}{\Delta y_k}\dd t_k
\end{multline}
where $i_0\coloneqq 0$, $i_{K+1}\coloneqq N$, $t_0\coloneqq 0$, $t_{K+1}\coloneqq 1$ and
\begin{equation}
\Delta t_k\coloneqq t_{k}-t_{k-1},\quad \Delta i_k\coloneqq i_{k}-i_{k-1}.
\end{equation}
Note that the previous equation has the form of a multinomial distribution. Introducing the rescaled variable $m_k(t_k)=\sqrt{N}\left(t_k-\frac{i_k}{N}\right)$, in the large $N$ limit we obtain a multivariate Gaussian distribution in the variable $\{\Delta m_k\}_{k=1,\dots,K}$, $\Delta m_k\coloneqq m_k-m_{k-1}$, $m_0\coloneqq 0$, whose covariance matrix is given by the degenerate matrix
\begin{equation}
\mathsf \Sigma_{ij}(\{y\})=\delta_{ij} \Delta t_i-\Delta t_i\Delta t_j.
\end{equation}
Note that $\det\mathsf \Sigma=\left(1-\sum_{k=1}^{K+1}\Delta t_k\right)\prod_{k=1}^{K+1}\Delta t_k=0$: the constraint $\sum_{i}\Delta t_i=1$ reduces the rank of the matrix from $K+1$ to $K$ (we have indeed $K$ independent variables). In this case, no density distribution exists in the $(K+1)$-dimensional space of variables $\Delta m_1,\dots,\Delta m_{K+1}$, due to the fact that an additional constraint, $\sum_{k=1}^{K+1}\Delta m_k=0$, holds, and therefore the distribution is \textit{singular}. We need to restrict therefore to the subspace of $K$ variables $\{\Delta m_k\}_{k=1,\dots K}$. The distribution is still a multivariate Gaussian but with covariance matrix $\Sigma_{ij}=\mathsf \Sigma_{ij}$ with $i,j=1,\dots,K$. The covariance matrix is positive definite and non singular on this subspace, being $\det \Sigma=\left(1-\sum_{k=1}^{K}\Delta t_k\right)\prod_{k=1}^{K}\Delta t_k=\prod_{k=1}^{K+1}\Delta t_k$ and
\begin{equation}
\left[\Sigma(y)\right]_{ij}^{-1}=\frac{\delta_{ij}}{\Delta t_i}+\frac{1}{\Delta t_{K+1}}
\end{equation}
The joint distribution is therefore
\begin{multline}\label{JBB}
\Pr\left(\{m_k(t_k)\in \dd m_k\}_{k=1,\dots K}\right)=\\
=\frac{\prod_{k=1}^K\dd m_k}{\sqrt{(2\pi)^K\prod_{k=1}^{K+1}\Delta t_k}}\exp\left(-\sum_{k=1}^K\frac{(\Delta m_k)^2}{2\Delta t_k}-\frac{m_K^2}{2\Delta t_{K+1}}\right).
\end{multline}
The previous distribution is exactly the joint distribution for the Brownian bridge process, proving the equivalence of the processes in the large $N$ limit. In particular, the two point joint distribution for the Brownian bridge is given by eq.~\eqref{JBB} for $K=2$:
\begin{multline}
\Pr\left(m_1(t_1)\in \dd m_1,m_2(t_2)\in \dd m_2\right)=\\
=\frac{\dd m_1\dd m_2}{2\pi\sqrt{(1-t_2)(t_2-t_1)t_1}}\e^{-\frac{m_1^2}{2t_1}-\frac{(m_2-m_1)^2}{2(t_2-t_1)}-\frac{m_2^2}{2(1-t_2)}}\label{JBB2}
\end{multline}
where $0< t_1<t_2< 1$ is assumed.

\section{Probability distributions for the Brownian bridge}
In the present section we briefly present, without proofs, some fundamental probability results on the Brownian bridge process and on some noteworthy probability distributions related to it. As explained before, the distribution for the sup of the absolute value of a Brownian bridge is given by the well known Kolmogorov distribution, eq.~\eqref{kolmogorov}. Using the reflection principle and Bayes' theorems, it can be also proved that \citep{Beghin1999,dudley2002real}
\begin{multline}\textstyle
\Pr\left(\sup_{t\in[0,1]}\mathsf B(t)<M,\inf_{t\in[0,1]}\mathsf B(t)>-m\right)=\\\textstyle=1+2\sum_{k=1}^\infty\e^{-2k^2(M+m)^2}+\\\textstyle
-\sum_{k=0}^\infty\left(\e^{-2(km+(k+1)M)^2}+\e^{-2(kM+(k+1)m)^2}\right).\label{bbjmb}
\end{multline}

For $t\in(0,1)$ and $M>b$, it can be easily obtained that
\begin{multline}\textstyle
\Pr\left(\sup_{s\in[0,1]}\mathsf B(s)<M,\mathsf B(t)\in \dd b\right)=\\
\textstyle=\frac{\e^{-\frac{b^2}{2t(1-t)}}\left(1-\e^{-\frac{2M(M-b)}{t}}\right)\left(1-\e^{-\frac{2M(M-b)}{1-t}}\right)}{\sqrt{2\pi t(1-t)}}\dd b.\label{bbsup}
\end{multline}
The computation of \eqref{mepbc} can be easily performed using the distribution of the sup of the Brownian bridge
\begin{equation}
\Pr\left(\sup_{t\in[0,1]}\mathsf B(t)<M\right)=1-\e^{-2 M^2}.
\end{equation}
\bibliography{Biblio.bib}
\end{document}